\documentclass[aps,prl,longtable,superscriptaddress,showpacs,preprintnumbers,amsmath,amssymb]{revtex4}
\usepackage{longtable}
\newcommand{\ovl}{\overline}
\newcommand{\open}{{<\kern -0.3 em{\scriptscriptstyle )}}}
\usepackage{graphicx} % Include figure files
\usepackage{dcolumn}  % Align table columns on decimal point

\usepackage{epic,eepic}
\usepackage{fancybox}
\usepackage{psboxit}
\usepackage{axodraw}

\usepackage{pstricks,pst-node,pst-text}%,pst-osci}

\graphicspath{{ps}}

\begin{document}

%\vspace*{-3\baselineskip}
%\resizebox{!}{3cm}{\includegraphics{belle.eps}}

\preprint{%\vbox{ \hbox{ draft version 1.6  }
						\hbox{Belle DRAFT {\it 2011-5 }}
						\hbox{KEK Preprint {\it 2010-53
}}
%                        \hbox{Intended for {\it PRL}}
%                        \hbox{Author: R. Seidl, A. Vossen, M. Grosse-Perdekamp, A. Ogawa}
%                        \hbox{Committee: B. Golob(chair),W. Bartel, Y. Miyachi, }}
}
\title{ \quad\\[1.0cm] Observation of transverse polarization asymmetries of charged pion pairs in \boldmath$e^+e^-$\unboldmath annihilation near \boldmath $ \sqrt{s}=10.58 $ \unboldmath GeV.}

%%%% >>>>> insert the authorlist here. BEFORE the abstract !!!!! <<<<<
%%%% >>>>> from the authorship confirmation web page
%%% Name the file author.tex and use \input{author} to insert into your latex file.
%\author{Author}\affiliation{affiliation}
%%% Paper:    Interference fragmentation function for pi+ pi- 
%%% Journal:  Physical Review Letters
%%% Contacts: R. Seidl (rseidl@rcf.rhic.bnl.gov)
%%%           A. Vossen (vossen@uiuc.edu)
%%%           M. Grosse-Perdekamp (mgp@uiuc.edu)
%%%           A. Ogawa (ogawa@bnl.gov)
%%% Non-responding authors or those who said NO are commented out.
%%% ====================================================================
%%% Click the RELOAD button on your web browser to see the updated file.
%%% ====================================================================
%%% Use \input{author} to insert this material into your latex file.
%%%%% Force institutions to appear in alphabetical order when typeset.
\affiliation{Budker Institute of Nuclear Physics, Novosibirsk}
%%%\affiliation{Faculty of Mathematics and Physics, Charles University, Prague}
%%%\affiliation{Chiba University, Chiba}
%%%\affiliation{University of Cincinnati, Cincinnati, Ohio 45221}
\affiliation{Department of Physics, Fu Jen Catholic University, Taipei}
\affiliation{Deutsches Elektronen--Synchrotron, Hamburg} 
\affiliation{Justus-Liebig-Universit\"at Gie\ss{}en, Gie\ss{}en}
\affiliation{Gifu University, Gifu}
%%%\affiliation{The Graduate University for Advanced Studies, Hayama}
%%%\affiliation{Gyeongsang National University, Chinju}
\affiliation{Hanyang University, Seoul}
\affiliation{University of Hawaii, Honolulu, Hawaii 96822}
\affiliation{High Energy Accelerator Research Organization (KEK), Tsukuba}
%%%\affiliation{Hiroshima Institute of Technology, Hiroshima}
\affiliation{University of Illinois at Urbana-Champaign, Urbana, Illinois 61801}
%%%\affiliation{Indian Institute of Technology Guwahati, Guwahati}
\affiliation{Institute of High Energy Physics, Chinese Academy of Sciences, Beijing}
\affiliation{Institute of High Energy Physics, Vienna}
\affiliation{Institute of High Energy Physics, Protvino}
%%%\affiliation{Institute of Mathematical Sciences, Chennai}
%%%\affiliation{INFN - Sezione di Torino, Torino}
\affiliation{Institute for Theoretical and Experimental Physics, Moscow}
\affiliation{J. Stefan Institute, Ljubljana}
\affiliation{Kanagawa University, Yokohama}
\affiliation{Institut f\"ur Experimentelle Kernphysik, Karlsruher Institut f\"ur Technologie, Karlsruhe}
\affiliation{Korea Institute of Science and Technology Information, Daejeon}
\affiliation{Korea University, Seoul}
%%%\affiliation{Kyoto University, Kyoto}
\affiliation{Kyungpook National University, Taegu}
\affiliation{\'Ecole Polytechnique F\'ed\'erale de Lausanne (EPFL), Lausanne}
\affiliation{Faculty of Mathematics and Physics, University of Ljubljana, Ljubljana}
\affiliation{University of Maribor, Maribor}
%%%\affiliation{Max-Planck-Institut f\"ur Physik, M\"unchen}
%%%\affiliation{University of Melbourne, School of Physics, Victoria 3010}
\affiliation{Nagoya University, Nagoya}
%%%\affiliation{Nara University of Education, Nara}
\affiliation{Nara Women's University, Nara}
\affiliation{National Central University, Chung-li}
\affiliation{National United University, Miao Li}
\affiliation{Department of Physics, National Taiwan University, Taipei}
%%%\affiliation{H. Niewodniczanski Institute of Nuclear Physics, Krakow}
\affiliation{Nippon Dental University, Niigata}
\affiliation{Niigata University, Niigata}
\affiliation{University of Nova Gorica, Nova Gorica}
\affiliation{Novosibirsk State University, Novosibirsk}
\affiliation{Osaka City University, Osaka}
%%%\affiliation{Osaka University, Osaka}
%%%\affiliation{Panjab University, Chandigarh}
%%%\affiliation{Peking University, Beijing}
%%%\affiliation{Princeton University, Princeton, New Jersey 08544}
\affiliation{Research Center for Nuclear Physics, Osaka}
\affiliation{RIKEN BNL Research Center, Upton, New York 11973}
\affiliation{Saga University, Saga}
\affiliation{University of Science and Technology of China, Hefei}
\affiliation{Seoul National University, Seoul}
%%%\affiliation{Shinshu University, Nagano}
\affiliation{Sungkyunkwan University, Suwon}
\affiliation{School of Physics, University of Sydney, NSW 2006}
\affiliation{Tata Institute of Fundamental Research, Mumbai}
%%%\affiliation{Excellence Cluster Universe, Technische Universit\"at M\"unchen, Garching}
%%%\affiliation{Toho University, Funabashi}
\affiliation{Tohoku Gakuin University, Tagajo}
\affiliation{Tohoku University, Sendai}
\affiliation{Department of Physics, University of Tokyo, Tokyo}
\affiliation{Tokyo Institute of Technology, Tokyo}
\affiliation{Tokyo Metropolitan University, Tokyo}
\affiliation{Tokyo University of Agriculture and Technology, Tokyo}
%%%\affiliation{Toyama National College of Maritime Technology, Toyama}
\affiliation{CNP, Virginia Polytechnic Institute and State University, Blacksburg, Virginia 24061}
\affiliation{Wayne State University, Detroit, Michigan 48202}
%%%\affiliation{Yamagata University, Yamagata}
\affiliation{Yonsei University, Seoul}
  \author{A.~Vossen}\affiliation{University of Illinois at Urbana-Champaign, Urbana, Illinois 61801}\affiliation{Indiana University, Bloomington, Indiana 47408} % UIUC
  \author{R.~Seidl}\affiliation{RIKEN BNL Research Center, Upton, New York 11973} % RIKEN
  \author{I.~Adachi}\affiliation{High Energy Accelerator Research Organization (KEK), Tsukuba} % KEK
% \author{K.~Adamczyk}\affiliation{H. Niewodniczanski Institute of Nuclear Physics, Krakow} % Krakow
  \author{H.~Aihara}\affiliation{Department of Physics, University of Tokyo, Tokyo} % Tokyo
% \author{K.~Arinstein}\affiliation{Budker Institute of Nuclear Physics, Novosibirsk}\affiliation{Novosibirsk State University, Novosibirsk} % BINP
% \author{Y.~Arita}\affiliation{Nagoya University, Nagoya} % Nagoya
% \author{T.~Aso}\affiliation{Toyama National College of Maritime Technology, Toyama} % Toyama
% \author{V.~Aulchenko}\affiliation{Budker Institute of Nuclear Physics, Novosibirsk}\affiliation{Novosibirsk State University, Novosibirsk} % BINP
  \author{T.~Aushev}\affiliation{\'Ecole Polytechnique F\'ed\'erale de Lausanne (EPFL), Lausanne}\affiliation{Institute for Theoretical and Experimental Physics, Moscow} % ITEP
% \author{T.~Aziz}\affiliation{Tata Institute of Fundamental Research, Mumbai} % Tata
% \author{A.~M.~Bakich}\affiliation{School of Physics, University of Sydney, NSW 2006} % Sydney
  \author{V.~Balagura}\affiliation{Institute for Theoretical and Experimental Physics, Moscow} % ITEP
% \author{Y.~Ban}\affiliation{Peking University, Beijing} % Peking
% \author{E.~Barberio}\affiliation{University of Melbourne, School of Physics, Victoria 3010} % Melbourne
\author{W.~Bartel}\affiliation{Deutsches Elektronen--Synchrotron, Hamburg} % DESY
% \author{A.~Bay}\affiliation{\'Ecole Polytechnique F\'ed\'erale de Lausanne (EPFL), Lausanne} % Lausanne
% \author{I.~Bedny}\affiliation{Budker Institute of Nuclear Physics, Novosibirsk}\affiliation{Novosibirsk State University, Novosibirsk} % BINP
% \author{K.~Belous}\affiliation{Institute of High Energy Physics, Protvino} % Protvino
% \author{V.~Bhardwaj}\affiliation{Panjab University, Chandigarh} % Panjab
% \author{B.~Bhuyan}\affiliation{Indian Institute of Technology Guwahati, Guwahati} % IITG
  \author{M.~Bischofberger}\affiliation{Nara Women's University, Nara} % Nara
% \author{S.~Blyth}\affiliation{National United University, Miao Li} % NUU
  \author{A.~Bondar}\affiliation{Budker Institute of Nuclear Physics, Novosibirsk}\affiliation{Novosibirsk State University, Novosibirsk} % BINP
% \author{G.~Bonvicini}\affiliation{Wayne State University, Detroit, Michigan 48202} % WayneState
% \author{A.~Bozek}\affiliation{H. Niewodniczanski Institute of Nuclear Physics, Krakow} % Krakow
  \author{M.~Bra\v{c}ko}\affiliation{University of Maribor, Maribor}\affiliation{J. Stefan Institute, Ljubljana} % Ljubljana
% \author{J.~Brodzicka}\affiliation{H. Niewodniczanski Institute of Nuclear Physics, Krakow} % Krakow
% \author{O.~Brovchenko}\affiliation{Institut f\"ur Experimentelle Kernphysik, Karlsruher Institut f\"ur Technologie, Karlsruhe} % Karlsruhe
% ...
% BINP
% \author{T.~Zivko}\affiliation{J. Stefan Institute, Ljubljana} % Ljubljana
% \author{A.~Zupanc}\affiliation{Institut f\"ur Experimentelle Kernphysik, Karlsruher Institut f\"ur Technologie, Karlsruhe} % Karlsruhe
% \author{N.~Zwahlen}\affiliation{\'Ecole Polytechnique F\'ed\'erale de Lausanne (EPFL), Lausanne} % Lausanne
% \author{O.~Zyukova}\affiliation{Budker Institute of Nuclear Physics, Novosibirsk}\affiliation{Novosibirsk State University, Novosibirsk} % BINP
\collaboration{The Belle Collaboration}
\noaffiliation
%% end author list

\begin{abstract}
%\linenumbers
The interference fragmentation function translates the fragmentation of a quark with a transverse projection of the spin into an azimuthal asymmetry of two final-state hadrons. %Just like the Collins fragmentation function it is a chiral-odd fragmentation function which can be used to measure the quark transversity distribution in transversely polarized semi-inclusive DIS and pp experiments, however it does not depend on unintegrated transverse momentum. 
In $e^+e^-$ annihilation the product of two interference fragmentation functions is measured. %Since only small acceptance effects are observed it is possible to extract the azimuthal asymmetries in fits to the normalized yields. 
We report nonzero asymmetries for pairs of charge-ordered $\pi^+\pi^-$ pairs, which indicate a significant interference fragmentation function in this channel. 
The results are obtained from a $\, 672$ fb$^{-1}$ data sample that contains $711 \times 10^6$ $\pi^+\pi^-$ pairs and was collected at and near the $\Upsilon(4S)$ resonance, with the Belle detector at the KEKB asymmetric-energy $e^+ e^-$ collider.
%{\it For official integrated luminosity, see }
%{\tt     http://belle.kek.jp/group/ecl/private/lum/, }
%{\it  and for number of $B\bar{B}$ pairs, see }
%{\tt     http://belle.kek.jp/secured/nbb/nbb.html.   }
\end{abstract}


\maketitle

%%%% >>>> keep the final version single-spaced
%\tighten

{\renewcommand{\thefootnote}{\fnsymbol{footnote}}}
\setcounter{footnote}{0}

%\linenumbers
The transverse spin structure of the nucleon is only poorly understood as its extraction requires the knowledge of spin-dependent fragmentation functions. Here we report the observation of transverse asymmetries of charged pion pairs in $e^+ e^-$ annihilation near a center of mass energy of 10.58 GeV. These results can be used to extract the interference fragmentation function (IFF). 

The IFF, first suggested by Collins \cite{Collins:1993kq}, is sensitive to the transverse polarization of the fragmenting quark and thus can be used as a quark polarimeter. %Its measurement will be an important input to the analysis of quark transversity from asymmetries measured in semi inclusive deep inelastic scattering and polarized proton proton collisions.
The previous measurement of the Collins fragmentation function \cite{belleprl,Collinsprd} with the Belle detector allowed the first global analysis of transversity \cite{alexei} to be performed using data from HERMES \cite{hermescollins} and COMPASS \cite{compassd}.
Knowledge of the IFF will allow complementary access to transversity and a comparison to the Lattice QCD calculations \cite{Gockeler:2005aw}.
Moreover, by detecting a second hadron, the sensitivity to the quark spin
survives integration over transverse momenta. 
Thus, unlike the Collins effect, collinear models can be used 
for factorization and the QCD evolution of the fragmentation function is known \cite{Ceccopieri:2007ip}.
Like the Collins function, the IFF is chiral-odd and can be used to extract transversity from asymmetries measured in polarized semi-inclusive deep inelastic scattering (SIDIS) \cite{hermesiff,compassiff} or proton-proton scattering \cite{yangiff}.
%Therefore knowing $H_1^\sphericalangle$ opens up an independent way to measure transversity with some attractive theoretical features.

\begin{figure}[th]
\vspace*{1cm}
\begin{center}
\includegraphics[width=0.45\textwidth]{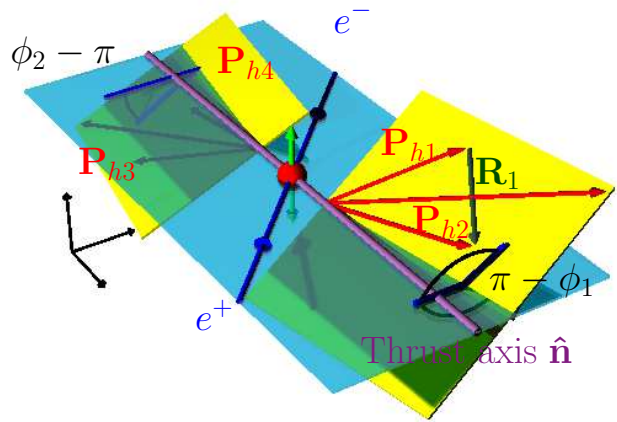}
%\includegraphics[width=0.45\textwidth]{plots/plotangle2.eps}
%\caption{\label{fig:angles} Azimuthal angle definitions for the two methods. Left plot: Azimuthal angles $\phi_1$ and $\phi_2$ as defined relative to the thrust axis. Right plot:  Azimuthal angles $\phi_{1R}$ and $\phi_{2R}$ as defined relative to the hadron pairs momentum axis $\mathbf{P}_3+\mathbf{P}_4$ of the second hemisphere.}
\caption{\label{fig:angles} Azimuthal angle definitions for $\phi_1$ and $\phi_2$ as defined relative to the thrust axis in the CMS.}
\end{center}
\end{figure}

The quantity sensitive to the transverse polarization of quarks is a cosine modulation of the azimuthal angle $\phi$ of the plane spanned by the momenta of the two hadrons ${h}_1$, ${h}_2$
around the fragmenting quark direction with respect to the transverse quark spin.
However, while the quark spin is unknown in unpolarized $e^+$ $e^-$ scattering, the two primordial quarks appear in two back-to-back jets. The kinematics of the process is shown in Fig.~\ref{fig:angles}. Thus, instead of measuring the azimuthal angle between 
the spin vector and the vector $\mathbf{R}=\mathbf{P}_{h1}-\mathbf{P}_{h2}$ describing the two-hadron-plane, one measures an azimuthal correlation of two hadron pairs detected in opposite hemispheres $\alpha={1,2}$.  The angles
$\phi_1$ and $\phi_2$ are defined in the center-of-mass system (CMS) between $\mathbf{R}_{\alpha}$ and the event plane spanned by the electron-positron axis $\hat{\mathbf{z}}$ and the thrust axis $\hat{\mathbf{n}}$ \cite{Artruiff}. %Normalizing the two hadrons' momenta by their momentum as initially suggested \cite{Artruiff} has little impact on the corresponding angles and therefore has no impact on the results.  
%Weighting the $\cos(\phi_1+\phi_2)$ amplitude with the kinematical factor $\frac{1}{4}\cos(\theta)$, the forwardness of the process,
%then selects events in which the spin projection of the virtual photon on the electron-positron axis is 0, and thus the quark
%spins are transverse to the quark-antiquark axis and antiparallel.
They can be expressed in terms of measured quantities as:
\begin{eqnarray}
\phi_{\{1,2\}}&=&\mathrm{sgn}\left[\hat{\mathbf{n}}\cdot (\hat{\mathbf{z}} \times \hat{\mathbf{n}} \times (\hat{\mathbf{n}}\times \mathbf{R}_{1,2})\}\right] \nonumber \\
&\times &\arccos \left(\frac{\hat{\mathbf{z}}\times \hat{\mathbf{n}}}{|\hat{\mathbf{z}}\times \hat{\mathbf{n}}|}
\cdot \frac{\hat{\mathbf{n}}\times \mathbf{R}_{1,2}}{|\hat{\mathbf{n}}\times \mathbf{R}_{1,2}|}
\right)\quad .
\end{eqnarray}
As in the Collins analysis, a second method can be applied, which does not directly depend on the thrust axis to calculate the angles, but defines the reference axis via the momentum of the second hadron pair and corresponding angles $\phi_{1R}$ and $\phi_{2R}$.
Using either set of angles, $\phi_1,\phi_2$ or $\phi_{1R},\phi_{2R}$, one can obtain a $\cos(\phi_{1(R)}+\phi_{2(R)})$ modulation proportional to the interference fragmentation functions normalized by the corresponding unpolarized di-hadron fragmentation functions. 
The amplitude of this modulation in $e^+e^-$ annihilation is according to Boer \cite{Boer:2003ya}:
\begin{widetext}
\begin{eqnarray}
a_{12R}(z_1, z_2, m_1^2, m_2^2) &\propto & \frac{1}{2} \; \frac{\sin^2\theta}{1+\cos^2\theta} \; \left[ 
\sum_{q, \ovl q}\; e_q^2 \;z_1^2 z_2^2\;  H_{1}^{\open \,q} (z_1, m_1^2) \; { H}_{1\,}^{\open \, \ovl q}
(z_2, m_2^2) \right] \nonumber \\
& &\times \left[ \sum_{q, \ovl q}\; e_q^2 \, z_1^2 \, z_2^2\; 
D_1^q (z_1, m_1^2) \; { D}_1^{\ovl q}(z_2, m_2^2) \right]^{-1} \quad,
\label{eq:asy}
\end{eqnarray}
\end{widetext}
and a similar formula for the $\cos(\phi_1+\phi_2)$ modulation amplitude $a_{12}$. %, which uses the thrust axis to define the reference axis. 
The interference fragmentation function $H^{\open,q}_1$ of a quark $q$ ( and charge $e_q$) , and its polarization-independent counterpart $D_1^q$, depend on the fractional energy $z_\alpha \stackrel{CMS}{=} 2E_\alpha/\sqrt{s}$ of the hadron pair in hemisphere $\alpha$ and on its invariant mass $m_\alpha$. The CMS energy is denoted by $\sqrt{s}$ and the polar angle $\theta$ is defined between the beam axis and the reference axis in the CMS. As dependence on the polar angule is a clear indication of initial transverse quark polarization, this dependence was studied. %A nonzero linear dependence on the polar angular term was seen in this analysis, however the constant term was also nonzero. % ividual masses $M_1, M_2$ of the 4 participating hadrons as well as the kinematic functions $A(y)= 1+\cos^2(\theta)$ and $B(y)=sin^2(\theta)$ appear in equation \ref{eq:asy}. 

Collins and Ladinsky\cite{linSigModel} used the linear sigma model to make the first predictions for $\pi$-$\pi$ correlations. % as a probe for the fragmentation of polarized quarks \cite{linSigModel}.
Another approach makes use of a partial wave analysis to arrive at predictions for $H_1^\sphericalangle$, which receives essential contributions from the interference of meson pairs (pions and kaons) in relative S- and P-wave states %The insights gained from the partial wave analysis inspired model predictions for $H_1^\sphericalangle$ based on the interference of meson pairs (pions and kaons) in relative s- and p-waves
\cite{Bacchetta:2002ux,accessingTransvWIFF,Bianconi:1999cd}.
A strong dependence on the invariant mass of the hadron pair is predicted. %, because the relative partial waves of the meson pairs are determined by the resonance they were produced from.
%Exemplarily for predictions of effects at Belle kinematics are the models by 
Predictions for spin effects that can be observed at the B-factories can be found in papers by Jaffe, Jin and Tang \cite{phaseShiftModel} and from references \cite{twoHadIFFModCalc,modDiHadFF}, with the latter being recently extended to $e^+e^-$ annihilation \cite{Radici} at Belle energies.
Jaffe and collaborators estimate the final-state interactions of the meson pairs from meson-meson phase shift data in \cite{phaseShiftData}, where it is observed that S- and P-wave production channels interfere strongly in the mass region around the $\rho$, the $K^*$ and the $\phi$ meson resonances, and give rise to a sign change of the IFF.

This analysis is based on a $\, 672$ fb$^{-1}$ data sample
collected  with the Belle detector at the KEKB asymmetric-energy
$e^+e^-$ (3.5 on 8~GeV) collider~\cite{KEKB}
operating at the $\Upsilon(4S)$ resonance and $60$ MeV below.
The Belle detector is a large-solid-angle magnetic
spectrometer that consists of a silicon vertex detector (SVD),
a 50-layer central drift chamber (CDC), an array of
aerogel threshold Cherenkov counters (ACC),  % <- \v{C}erenkov 2007.08
a barrel-like arrangement of time-of-flight
scintillation counters (TOF), and an electromagnetic calorimeter (ECL)
comprised of CsI(Tl) crystals located inside 
a superconducting solenoid coil that provides a 1.5~T
magnetic field.  An iron flux-return yoke located outside of
the coil is instrumented to detect $K_L^0$ mesons and to identify
muons (KLM).  The detector is described in detail elsewhere~\cite{Belle}.
% {\bf SVD2+SVD1, up to experiment 37:}
%Two inner detector configurations were used. A 2.0 cm radius beampipe
%and a 3-layer silicon vertex detector were used for the first sample
%of 157 fb$^{-1}$, while a 1.5 cm radius beampipe, a 4-layer
%silicon detector and a small-cell inner drift chamber were used to record  
%the remaining 516 fb$^{-1}$\cite{svd2}.  

%\subsection{Event selection}
%As this analysis deals with effects produced from transversely polarized initial quarks both continuum and regular on\_resonance events were analyzed but treated separately for obtaining the asymmeties and are only combined as a last step. All on\_resonance and continuum data from experiments 7 to 55 have been analyzed yielding a luminosity of approximately 588 fb$^{-1}$ on\_resonance and 73 fb$^{-1}$ in the continuum. The detailed luminosities and amounts of files were summarized in Table \ref{tab:datastat}. Being interested in hadronic events created from quark-antiquark pairs the HadronB{J} skim was used.
The most important selection criterion is the event shape variable thrust, $T$, the maximum of which defines the thrust axis $\hat{\mathbf{n}}$ :
\begin{equation}
T \stackrel{max}{=} \frac{ \sum_h|\mathbf{P^{\mathrm{CMS}}_h}\cdot\mathbf{\hat{n}}|}{ \sum_h|\mathbf{P^{\mathrm{CMS}}_h}|}\quad.
\end{equation}
The sum extends over all detected particles, and $P_h^{CMS}$ denotes their momenta in the CMS. The cosine of the deviation from reconstructed thrust axis and generated quark-antiquark pair axis for light quarks is 0.990 with an RMS of 0.015, as obtained from the simulated sample of events using the PYTHIA \cite{Sjostrand:1995iq} event generator and a GEANT \cite{geant} detector simulation. This value is compatible with those cited earlier in the Collins analysis \cite{belleprl}. %The thrust value varies between 0.5 for spherical events and 1 where all tracks are aligned with the thrust axis of an event. 
Since the two pairs of hadrons should appear in a two-jet topology, events are selected with a thrust value larger than 0.8. The contamination from $B$ decays in this event sample is around 2\% %and are generally found at the lower fractional energies z of the hadron pairs 
\cite{Collinsprd}. 
As the hadron pairs are sampled only in the barrel region of the detector, one has to ensure that for those pairs all possible azimuthal angles around the thrust axis lie also within this acceptance. For this purpose only events with a thrust axis pointing into the central detector are considered with the $z$ component of the thrust unit vector %(and thus the cosine of the polar angle of the thrust axis) to be 
$|\hat{\mathbf{n}}_z| < 0.75$.
%\begin{equation}
%|\hat{\mathbf{n}_z}| < 0.75 \quad.
%\end{equation}
In order to obtain a reliable thrust axis and to reduce the contribution from $e^+e^-\rightarrow\tau^+\tau^- $ events, the reconstructed energy of an event is required to be above 7~GeV. Tracks are required to lie in the central part of the detector acceptance corresponding to $ -0.6  < \cos(\theta_{LAB}) < 0.9 $, where $\theta_{LAB}$ is the polar angle in the laboratory frame. This corresponds to a nearly symmetric track selection in the CMS frame, with the polar angle range $-0.79< \cos(\theta_{CMS}) < 0.74$.
%The vertex positions of the tracks were corrected with the interaction point positions and have to be originating in
All tracks are required to originate from a region around the reconstructed interaction point, which is defined by the requirements $ dr < 2$ cm and $ |dz| < 4$ cm,
where $dr$ and $dz$ are the distance of closest approach to the interaction point in the plane perpendicular to the beam direction and along the direction of the beams. 
Pions were selected among the reconstructed charged tracks by vetoing identified muons, electrons and protons, and requiring a kaon - pion particle identification likelihood to be larger than 0.7 \cite{TaggingNIM}. With these requirements the fraction of fake pions in the selected sample is between 2.7 and 3.3\%. The overall fraction of misidentified pions, obtained from simulated data, is added as a relative systematic uncertainty of the final measured asymmetries and is correlated between the bins defined below. 
All pions are required to have a minimal fractional energy $z=\frac{2E_{h}}{\sqrt{s}}>0.1 $. The fractional energy $z_\alpha$ of each pion pair is thus at least 0.2. The effect of the minimal hadron energy requirement on the decay angular distribution will be discussed later.

In addition to $\theta_{LAB}$, other polar angles in this analysis are the polar angle of the thrust axis in the CMS $\theta_t=\mathrm{acos}(\hat{\mathbf{n}}_z)$ and the decay angles of a hadron pair in their respective center-of-mass systems $\theta_{1d,2d}$ defined with respect to the first (i.e., positive) hadron. The lowest-order interference fragmentation term has a $\sin \theta_{d}$ distribution.

Any combination of two charged pions with opposite charge is combined in a pair if the two hadrons are in the same hemisphere. %by requiring $r_{hemi,ij}>0$, where $r_{hemi}$ is defined as:
%$r_{hemi}=(\mathbf{P}_{i}\cdot \hat{\mathbf{n}})(\mathbf{P}_{j}\cdot \hat{\mathbf{n}})$, 
%where $\hat{\mathbf{n}}$ is the unit-vector of the thrust axis and $\mathbf{P_{\i,j}}$ are the momenta of the individual tracks in the CMS frame. %Two pairs of pion pairs are further considered for this analysis if the two pairs appear in
For the analysis we select two pion pairs belonging to opposite hemispheres. % by requiring $r_{hemi,\alpha\beta}<0$.
In addition, the requirement of an opening angle relative to the thrust axis $\cos\psi=|(\hat{\mathbf{n}}\cdot \mathbf{P}_{h})|/|\mathbf{P}_{h}|>0.8$ selects only tracks that have at least a certain fraction of their momentum along the thrust axis. After these selection criteria, the total data sample contains $711 \times 10^6$ $\pi^+\pi^-$ pairs (1.58 di-pion pairs per event).
Throughout this paper the order of the pion pairs used for calculating $\mathbf{R}_{1,2}$ is always $\pi^+\pi^-$ in both hemispheres. The data is binned in either $8\times8$ $m_1,m_2$ bins between $0.25$ GeV/$c^2$ and $2$ GeV/$c^2$ or in $9\times 9$ $z_1, z_2$ bins between $0.2$ and $1.0$.
The first method of assessing the interference fragmentation function is based on measuring a $\cos(\phi_1+\phi_2) $ modulation of two hadron pair yields ($N(\phi_1+\phi_2)$) on top of the flat distribution due to the unpolarized part of the fragmentation functions. The unpolarized part is given by the average bin content $\langle N_{12}\rangle$. The normalized distribution is then defined as
 \begin{equation}
R_{12}:=\frac{N(\phi_1+\phi_2)}{\langle N_{12}\rangle} \quad.
\label{eq:r12def}
\end{equation} 
The two-pion pair yields $N(\phi_{1(R)}+\phi_{2(R)})$ are obtained for each kinematic bin in 16 equal-size bins of the azimuthal angles. %The two asymmetry reconstruction methods with respect to the reference axes introduced earlier use the angles $\phi_1+\phi_2$ and $\phi_{1R}+\phi_{2R}$.
The normalized azimuthal di-hadron yields, $R_{12(R)}$ can be parameterized as:
\begin{eqnarray}
R_{12(R)} =a_{12(R)} \cos(\phi_{1(R)}+\phi_{2(R)}) + b_{12(R)} + \quad&& \nonumber \\ 
   c_{12(R)} \sin (\phi_{1(R)}+\phi_{2(R)}) + d_{12(R)}\cos2(\phi_{1(R)}+\phi_{2(R)})&&
\label{eq:yieldfit}
\end{eqnarray}
where the parameter $b_{12(R)}$ should be unity due to the normalization. The parameter $a_{12(R)}$ is the amplitude proportional to the interference fragmentation functions. The normalized distribution is fit to equation (\ref{eq:yieldfit}) with $a_{12(R)}$, $b_{12(R)}$, $c_{12(R)}$ and $d_{12(R)}$ as free parameters. The reduced $\chi^2$ values of the individual fits over all run ranges and bins are well described by a $\chi^2$ distribution with a mean value close to unity.  %by a least $\chi^2$ fit to obtain these values. 
%In addition one can also include higher order modulations in the fit which ideally should not interfere with $a_{12(R)}$. 

The PYTHIA event generator used in this analysis does not contain the spin effects related to the IFF, and thus all asymmetries are expected to vanish. A check is performed for the kinematic effects that could mimic the spin-induced asymmetries. For this purpose light quark (uds) events and charm quark events have been generated, which were tracked through the detector in a GEANT simulation and then fully reconstructed. Asymmetries were evaluated at the generated 4-momentum level, as well as for reconstructed events.  The results of this analysis are summarized in Table \ref{tab:mcvalues}, where effects of a finite detector acceptance are clearly visible. They can be significantly reduced via the opening angle selection.
%The asymmetries for Pythia \cite{Sjostrand:1995iq} generated uds-quark-antiquark MC in $4\pi$ acceptance, in the detector acceptance, fully tracked uds MC as well as charm MC are summarized in table \ref{tab:mcvalues}. A nonzero acceptance effect can be seen for both reconstruction methods when confined to the detector acceptance which can be eliminated with the opening angle selection mentioned above. 
The sum of the absolute value of the reconstructed asymmetries and their statistical uncertainties in the simulated sample were assigned as bin-by-bin systematic uncertainties of the data asymmetries. They represent the largest systematic uncertainties, which are up to several \% in the lowest statistics bins.  
%\begin{figure}
%\begin{center}
%\includegraphics[width=0.45\textwidth]{exp-4_r3_asypanel_val0_mix500_m_comp0.eps}
%\caption{\label{fig:asymcacc}$a_{12}$ asymmetry parameters as a function of $m_2$ for the $m_1$ bins for $\pi^+\pi^-$ pairs from generated MC in the Belle acceptance without (blue triangles) and with an opening cut of 0.8 (red triangles). In addition the differences between both selection and a constant fit over all bins is displayed in the last panel.}    
%\end{center}
%\end{figure}

\begin{table}[th]
\caption{MC results in \% averaged over all $z$ bins for generated uds events (uds gen), within the geometrical acceptance (uds gen. acc.) as well as reconstructed uds and charm events.\label{tab:mcvalues}}
\begin{tabular}{c rr }\hline
Sample &\multicolumn{2}{c}{$z_1,z_2$-Asymmetries}\\
%& &\multicolumn{2}{c}{Method 1}&\multicolumn{2}{c|}{Method 2}\\
&\multicolumn{1}{c}{$\langle a_{12}\rangle$}&\multicolumn{1}{c}{$\langle a_{12R}\rangle$}\\ \hline
\multicolumn{3}{c}{No opening angle cut}\\ \hline
uds gen.& $-0.089\pm0.008$ &$-0.108\pm0.008$ \\
uds gen. acc. &$-0.488\pm0.011$ &$-0.490\pm0.011$\\
uds rec. & $-0.401\pm0.007$ &$-0.428\pm0.007$ \\
charm rec. & $-0.446\pm0.041$ &$-0.388\pm0.044$ \\ \hline
\multicolumn{3}{c}{With opening angle cut of 0.8}\\ \hline
uds gen.& $-0.038\pm0.013$ &$-0.035\pm0.013$ \\
uds gen. acc.& $-0.112\pm0.016$ &$-0.113\pm0.016$ \\
uds rec. &  $0.020\pm0.010$ &$0.006\pm0.010$ \\
charm rec. &  $0.006\pm0.040$ &$0.027\pm0.040$ \\
\hline
\end{tabular}
\end{table}

\begin{figure}[ht]
\begin{center}
\includegraphics[width=0.45\textwidth]{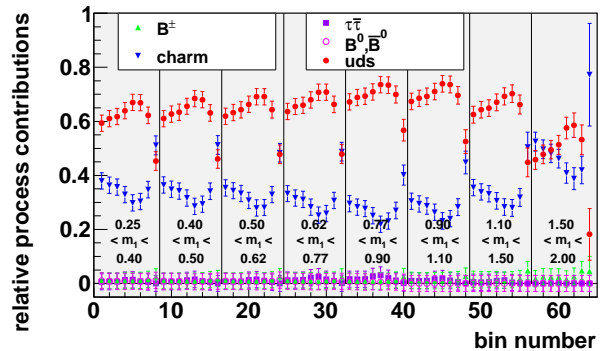}
\caption{Relative contributions of various processes for pion pairs as a function of the $8 \times 8$ $m_1, m_2$ bin number. The closed circles denote light quark-antiquark pair events, inverted triangles -- charm events, triangles -- charged $B$ meson pairs, open circles -- neutral $B$ meson pairs and  squares -- $\tau$ pairs. \label{fig:charmm}} 
\end{center}
\end{figure}

\paragraph{Mixed events:}
As the asymmetry requires a correlation between the hadron pairs on the quark and the antiquark side of an event, taking one hadron pair of another event should destroy this correlation and the asymmetries obtained for such a mixed-event data sample should vanish unless detector effects introduce artificial asymmetries. %have to be vanishing. Any possible dependence on the detector acceptance or a non homogeneous detector response can however be still existing in this mixed signal. 
Two ways of extracting event-mixed asymmetries were applied: using a hadron pair of a first event in combination with a pair of a second event, and taking the axis information either from the first or the second event. %For the method $a_{12}$ the thrust axis of that previous event was used as the reference axis, for method $a_{12R}$ the hadron pair momentum was used as the reference axis. . One mixing procedure, 
%The second mixing procedure, takes the axis information of the current event to determine the asymmetries. %The results for the event mixing can be seen in Fig.~\ref{fig:mixdata} for %generated MC with full acceptance, generated MC in the detector acceptance, reconstructed MC and data. No statistical significant deviations from zero can be seen. 
%A summary of different samples % and species%(generated uds MC, generated with acceptance cuts, reconstructed MC and data), averaged over the bins 
%can be found in Table \ref{tab:mixing} where only the largest asymmetry resulting from the two approaches to fix the reference axis is displayed. %for the larger of the two mixing methods. 
The values from data are  $(-0.019\pm0.017)$\% for $a_{12}$ and $(-0.012\pm0.017)$\% for $a_{12R}$. These values are included as absolute systematic uncertainties in the results. Studies of polarization build-up in the KEK rings were performed earlier and were consistend with no beam polarization \cite{Collinsprd}.
%The larger differences of the two methods to zero, which range from 0.1\% to 10\% in the final data have been assigned as systematic errors as summarized in tables \ref{tab:sytematicsm} to \ref{tab:sytematicsmz}. The large differences only appear in kinematic bins with the lowest statistics, such that one can conclude that they are statistics driven.   
%\begin{table}[ht]
%\caption{Mixing results averaged over the $z$ binning in. The results integrated over other binnings are nearly identical.\label{tab:mixing}}
%\begin{tabular}{c rr }\hline
%sample &\multicolumn{2}{c}{$z_1,z_2$-Asymmetries}\\
%%& &\multicolumn{2}{c}{Method 1}\\\hline
%&\multicolumn{1}{c}{$\langle a_{12}\rangle$}&\multicolumn{1}{c}{$\langle a_{12R}\rangle$}\\
%uds gen.& $0.00070\pm0.00013$ & $0.00030\pm0.00013$ \\
%uds gen. acc.& $0.00020\pm0.00016$ &$-0.00021\pm0.00016$ \\
%uds rec. &  $0.00038\pm0.00010$ &$0.00051\pm0.00010$ \\
%charm rec. & $-0.00024\pm0.00040$ &$-0.00017\pm0.00040$ \\
%Data &  $-0.00019\pm0.00017$ &$-0.00012\pm0.00017$ \\ \hline
%\end{tabular}
%\end{table}
\paragraph{Higher harmonics:}
The higher-order terms in Eq.~(\ref{eq:yieldfit}) are needed to reproduce the azimuthal variations well. Generally these different harmonics are orthogonal and should not interfere with each other, but a limited acceptance can introduce other asymmetries. The small differences in $a_{12(R)}$ of up to 1\% between either fitting the first two terms or all are assigned as a bin-by-bin systematic uncertainty.
\paragraph{Weighted MC asymmetries:}
Artificial asymmetries were introduced into the MC generator for hadron pairs around the quark-antiquark axis and then reconstructed to test the validity of the reconstruction method. %and the underestimation of the reconstructed asymmetries were tested. 
The $a_{12}$ asymmetries, which depend directly on using the thrust axis as a proxy for the quark-antiquark axis, are reconstructed to $(92\pm 1)$\% of the generated value, and the $a_{12R}$ asymmetries to $(99\pm 1)$\%. Corresponding correction factors are applied to the measured asymmetries %.The final asymmetries are corrected by the inverse of this factor 
and the uncertainties were assigned as a systematic error. 
\paragraph{Process contributions:} 
%In the HadronBJ data sample with the event and particle selection as described above the majority of events are light quark (uds quark-antiquark pair) events followed by charm quark-antiquark pairs. As shown in Belle notes \cite{Collins3} the 
The thrust selection alone already reduces the background from $\Upsilon(4S)$ decays to a negligible level. The charm contribution, however, has nearly the same thrust distribution as that for light quarks. % since the mass of charmed quarks leaves enough energy to the initial charmed mesons, that a two-jet like topology can be seen. 
On the other hand, since pions from charmed mesons are the product of a decay chain, the fractional energies %still have to decay at least once and mostly more than once to arrive at pion pairs these events
 fall off more rapidly than for light quarks. Therefore the relative charm contribution also falls off from nearly 50\% at lowest $z$ bins to a few \% at high $z$. % as a function of z. % as can be seen in Fig.~\ref{fig:charmz}. % is increasing with $z$ as in particular the low multiplicity hadronic tau decays can contribute there if the energy lost by the neutrino is small.     
The charm contribution in the mass bins first falls as can be seen in Fig.~\ref{fig:charmm} but then increases again for invariant masses around $1$ GeV/$\mathrm{c}^2$. %It is clear that this behviour cannot originate due to the small correlation between higher invariant mass and higher z as the charm contributions are falling off with z. A possible explanation could be two particles from early and late in the decay chain of charmed mesons which have a large relative opening angle between them and thus are at higher invariant masses. Obviously the highest mass bins are even dominated by charm contributions. 
%Since the asymmetries are sizable in the highest mass bins one has to conclude that %either the asymmetries due to light quarks are even larger to compensate for the large charm contributions without asymmetries or 
%the charm events contain a very similar asymmetry as the light quarks.  
%Analyzing the reconstructed $evtgen-charm$ on\_resonance MC events of experiments 39 and 41 shows no asymmetry directly due to the decays. The $5\times5$ mass binning split up in $2\times2$ $z$ bins shows some differences in the highest masses one has to conclude that there are also nonzero asymmetries originating from charm events.

%\begin{figure}[ht]
%\begin{center}
%\includegraphics[width=0.45\textwidth]{exp39_bg_z_bin2.eps}
%\caption{Relative contributions from top to bottom of light quark-antiquark pair events (red), charm events (green), charged B meson pairs (blue), neutral B meson pairs (yellow) and $\tau$ pairs (purple) as a function of $z_2$ for all $z_1$ bins in the $9 \times 9$ $z_1, z_2$ binning.\label{fig:charmz}} 
%\end{center}
%\end{figure}

\begin{figure}[ht]
\begin{center}
\includegraphics[width=0.5\textwidth]{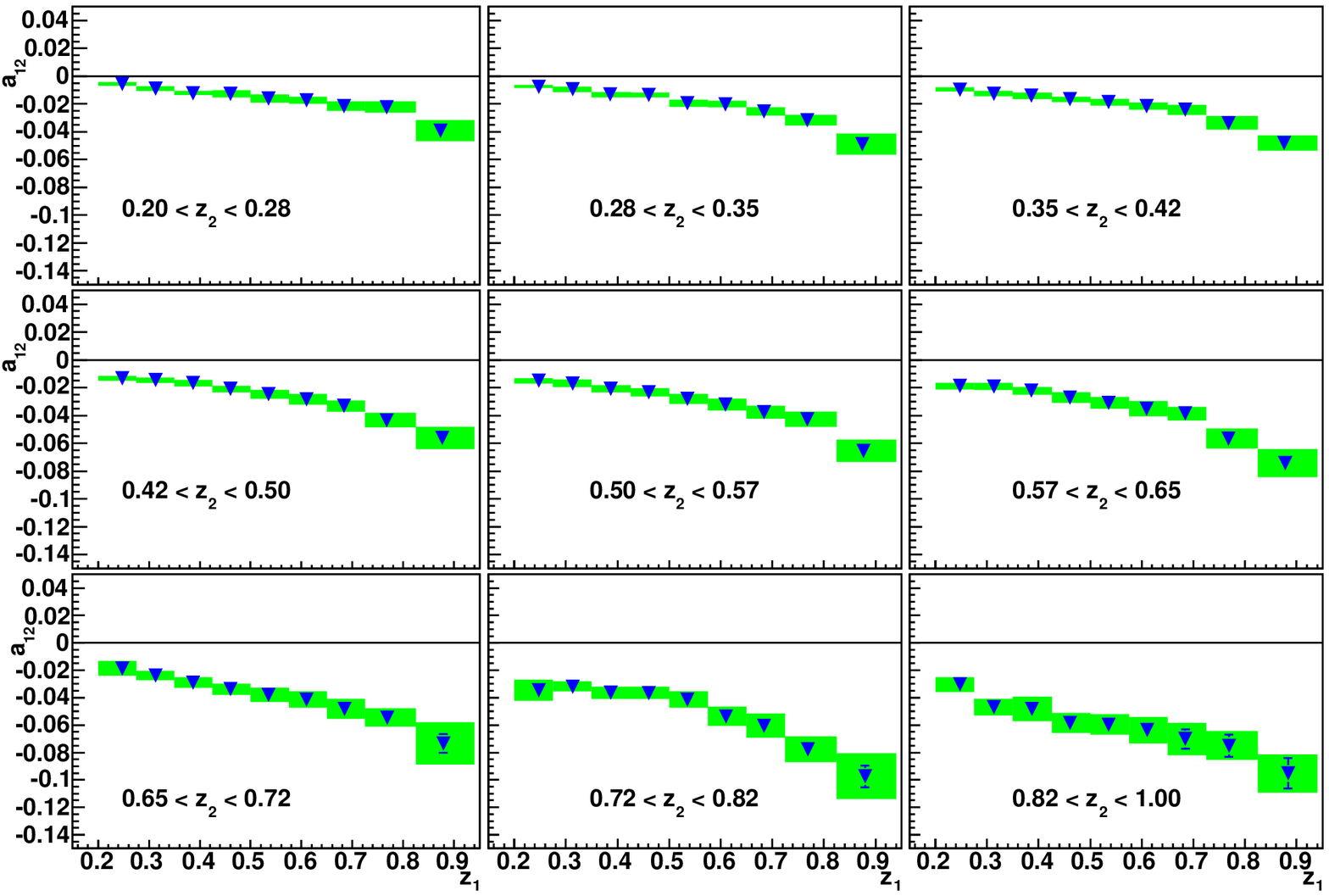}
\caption{\label{fig:finalzz}$a_{12}$ modulations for the $9\times9$ $z_1, z_2$ binning as a function of $z_1$ for the $z_2$ bins. The shaded (green) areas correspond to the systematic uncertainties.}
\end{center}
\end{figure}

\begin{figure}[ht]
\begin{center}
\includegraphics[width=0.5\textwidth]{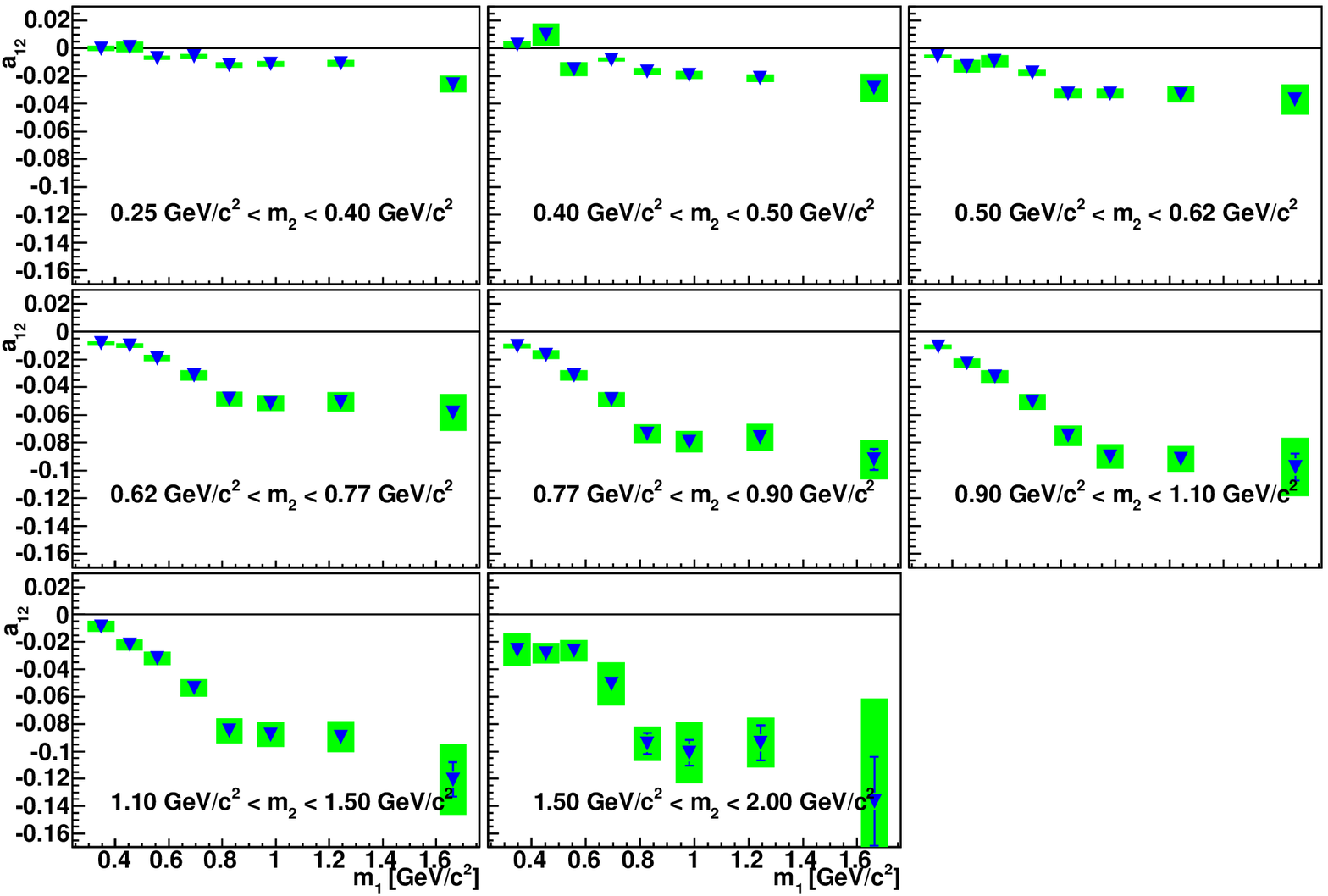}
\caption{\label{fig:finalmm}$a_{12}$ modulations for the $8\times8$ $m_1, m_2$ binning as a function of $m_2$ for the $m_1$ bins. The shaded (green) areas correspond to the systematic uncertainties.}
\end{center}
\end{figure}

There is a small contribution from $\tau$ pairs rising to several \% at high $z$.
When analyzing a $\tau$ enhanced data sample without the minimal energy requirement one finds asymmetries of $a_{12}=(-1.31\pm0.13)\%$ averaged over the whole kinematic range. This asymmetry can be explained by the sizeable residual contribution from continuum events in the $\tau$ enhanced data. %Generated and reconstructed asymmetries from $\tau?$ MC show that the effect intrinsic to the $\tau$ and its decays is compatible with zero. 
The relative contributions from $\tau$ pair events multiplied by their average asymmetry are added as systematic error, which is, however, negligibly small. %due to their small contribution the systematic errors are negligible.
\paragraph{Correlation studies:}
In order to exclude possible effects of correlations between different kinematic and azimuthal bins, MC studies have been performed, which did not find any such effects.
%In order to study possible effects due to the correlations amongst different kinematic and azimuthal bins reweighted generic MC data has been split into batches of 500k events. Asymmetries are evaluated for all batches %The whole analysis procedure to obtain the asymmetries has been applied to all these separated runs 
%using 5x5 $z$ and $m$ bins. The obtained asymmetries are then fit by a Gaussian to obtain the variation of the results. 
%The width of the Gaussian is compatible with the statistical error of the underlying sample and, thus, no systematic error is assigned due to possible correlations.
%%The Gaussian fit describes the result distribution quite well. The variation of the results is then compared to the average statistical error obtained in each of these samples and found to be consistent so that no systematic error has been assigned due to possible correlations.
%%
%%As seen in Fig.~\ref{fig:correlation} the with of the gaussian and the average statistical error are consistent so that no systematic error has been asigned 
%%due to possible correlations.
\paragraph{Inverted thrust selection:} The inverse thrust selection was also analyzed to test whether the azimuthal correlation of the two hadron pairs decreases. On average the asymmetries were 45\% smaller.
\paragraph{Results:}
The results can be seen in Fig.~\ref{fig:finalzz} as a function of the fractional energies and in Fig.~\ref{fig:finalmm} as a function of the di-pion invariant masses. One sees large asymmetries monotonically decreasing with fractional energy and invariant mass with an indication of leveling off at the highest invariant masses. %The overall magnitude is large; 
At higher masses or fractional energies an asymmetry of up to 10\% corresponds to interference fragmentation functions of more than 30\% the size of the corresponding unpolarized two-hadron fragmentation function.   
The results averaged over all kinematic bins are summarized in Table \ref{tab:final}%, corresponding to an observation of an effect at the 21$\sigma$ level
.  The $a_{12R}$ results show similar dependencies and magnitudes. 
All results, their central values and process fractions are tabulated in the electronic supplement to this publication which is appended in this preprint version.

\begin{table}
\begin{center}
\caption{Integrated asymmetries for the two reconstruction methods and their average kinematics.\label{tab:final}}
\begin{tabular}{c r}\\\hline
$\langle z_1 \rangle,\langle z_2 \rangle$ &0.4313\\
%$\langle z_2 \rangle$ &0.4312\\   
$\langle m_1 \rangle,\langle m_2 \rangle$ &0.6186 GeV/$c^2$ \\   
%$\langle m_2 \rangle$ &0.6186\\   
$\langle \sin^2\theta_t/(1+\cos^2\theta_t) \rangle$ &0.7636\\   
$\langle \sin\theta_{1d} \rangle,\langle \sin\theta_{2d} \rangle$ &0.9246\\   
%$\langle \sin\theta_{d2} \rangle$ &0.9246\\   
$\langle \cos\theta_{1d} \rangle,\langle \cos\theta_{2d} \rangle$ &0.0013\\   
%$\langle \cos\theta_{d2} \rangle$ &0.0014\\   
$a_{12}$&$ -0.0196 \pm  0.0002 (stat.) \pm  0.0022 (syst.)$\\
$a_{12R}$&$   -0.0179 \pm  0.0002 (stat.) \pm  0.0021 (syst.)$\\ \hline 
%$a_{12}$&$ -0.0199248 \pm  0.000188782 \pm  0.000879192$\\
%$a_{12R}$&$   -0.0177618 \pm  0.000175529(stat.) \pm  0.0007626(syst.)$\\ \hline 
\end{tabular}
\end{center}
\end{table}

\paragraph{Summary:} Large azimuthal asymmetries for two $\pi^+\pi^-$ pairs in opposite hemispheres were extracted from a 672 fb$^{-1}$ data sample. The asymmetries monotonically decrease as a function of $z_{1,2}$ and $m_{1,2}$ and no sign change is observed in contrast to \cite{phaseShiftModel}. The interference fragmentation function can be extracted from those asymmetries and used in a global fit to the SIDIS data \cite{hermesiff,compassiff} to obtain the transversity distribution function.

%***** Acknowledgments *****
We thank the KEKB group for excellent operation of the
accelerator, the KEK cryogenics group for efficient solenoid
operations, and the KEK computer group and
the NII for valuable computing and SINET3 network support.  
We acknowledge support from MEXT, JSPS and Nagoya's TLPRC (Japan);
ARC and DIISR (Australia); NSFC (China); 
DST (India); MEST, KOSEF, KRF (Korea); MNiSW (Poland); 
MES and RFAAE (Russia); ARRS (Slovenia); SNSF (Switzerland); 
NSC and MOE (Taiwan); NSF and DOE (USA).
\cleardoublepage
\subsection{Supplemental information for the paper: {\it  Observation of transverse polarization asymmetries of charged pion pairs in $e^+e^-$ annihilation near $\sqrt{s}=10.58$ GeV related to the interference fragmentation function.}}
The following contains the supplemental information available online together with the published journal version of this paper.
\begin{center}
\begin{longtable*}{rrrrrrrrr}
\caption{\label{tab:bgz} relative yields of uds, charm mixed, charged and tau contributions in the data sample for the integrated range $z_1,z_2 \in [0.2,1.0]$ and $m_1,m_2 \in [0.25,2.0]$ GeV/c$^2$.}\\
$\langle z_1 \rangle $ & $\langle z_2 \rangle $ & $\langle m_1 \rangle $ & $\langle m_2 \rangle $ & uds & charm & mixed & charged & $\tau$ \\ \hline
\endhead
\hline\multicolumn{9}{c}{continued on next page}
\endfoot
\hline \hline
\endlastfoot
$0.43$ &$0.43$ &$0.60$ &$0.60$ &$ 0.65\pm0.03$ &$ 0.32\pm0.03$ &$ 0.01\pm0.03$ &$ 0.01\pm0.03$ &$ 0.01\pm0.03$ \\ \hline
\end{longtable*}
\end{center}

\begin{center}
\begin{longtable*}{ccccc c  ccrr}
\caption{\label{tab:finalm} Final asymmetries and central values for the integrated data sample $z_1,z_2 \in [0.2,1.0]$ and $m_1,m_2 \in [0.25,2.0]$  GeV/c$^2$ in \%. The following abbreviations are used: $s^2 = \sin^2\theta_t$, $1+c^2=1+\cos^2\theta_t$, $s\theta_1=\sin \theta_1 $ and $s\theta_2= \sin \theta_2$. }\\
\hline $\langle z_1 \rangle $ & $\langle z_2 \rangle $ & $\langle m_1 \rangle $ & $\langle m_2 \rangle $ & $\langle s^2 \rangle/ \langle 1+c^2 \rangle $ & $\langle s\theta_1 \rangle $ & $\langle s\theta_2 \rangle$ &  $a_{12}$ & $a_{12R}$ \\ \hline
\endhead
\hline\multicolumn{9}{c}{continued on next page}
\endfoot
\hline \hline
\endlastfoot
$0.43$& $0.43$& $0.62$& $0.62$& $0.85/1.15$& $0.93$& $0.92$&   $-1.98\pm 0.02\pm 0.23$&   $-1.77\pm 0.02\pm 0.22$ \\
\end{longtable*}
\end{center}

\begin{center}
\begin{ruledtabular}
 \begin{longtable*}{rrrrrrrrr}
\caption{\label{tab:bgz} relative yields of uds, charm mixed, charged and tau contributions in the data sample for the the individual bins in the 9x9 $z_1\times z_2$ binning ($z_1,z_2 \in [0.2,0.275,0.35,0.425,0.5,0.575,0.65,0.725,0.825,1.0]$ and $m_1,m_2 \in [0.25,2.0]$ GeV/c$^2$).}\\
$\langle z_1 \rangle $ & $\langle z_2 \rangle $ & $\langle m_1 \rangle $ & $\langle m_2 \rangle $ & uds & charm & mixed & charged & $\tau$ \\  \hline
\endhead
\hline\multicolumn{9}{c}{continued on next page}
\endfoot
\hline \hline
\endlastfoot
$0.25$ &$0.25$ &$0.46$ &$0.46$ &$ 0.50\pm0.03$ &$ 0.46\pm0.03$ &$ 0.02\pm0.03$ &$ 0.02\pm0.03$ &$ 0.00\pm0.03$ \\ 
$0.25$ &$0.31$ &$0.46$ &$0.52$ &$ 0.52\pm0.03$ &$ 0.44\pm0.03$ &$ 0.02\pm0.03$ &$ 0.02\pm0.03$ &$ 0.00\pm0.03$ \\ 
$0.25$ &$0.39$ &$0.46$ &$0.58$ &$ 0.55\pm0.03$ &$ 0.42\pm0.03$ &$ 0.01\pm0.03$ &$ 0.02\pm0.03$ &$ 0.00\pm0.03$ \\ 
$0.25$ &$0.46$ &$0.46$ &$0.63$ &$ 0.60\pm0.03$ &$ 0.37\pm0.03$ &$ 0.01\pm0.03$ &$ 0.01\pm0.03$ &$ 0.00\pm0.03$ \\ 
$0.25$ &$0.53$ &$0.46$ &$0.68$ &$ 0.64\pm0.03$ &$ 0.33\pm0.03$ &$ 0.01\pm0.03$ &$ 0.01\pm0.03$ &$ 0.01\pm0.03$ \\ 
$0.25$ &$0.61$ &$0.46$ &$0.73$ &$ 0.69\pm0.03$ &$ 0.28\pm0.03$ &$ 0.01\pm0.03$ &$ 0.01\pm0.03$ &$ 0.01\pm0.03$ \\ 
$0.25$ &$0.68$ &$0.46$ &$0.77$ &$ 0.76\pm0.03$ &$ 0.21\pm0.04$ &$ 0.01\pm0.03$ &$ 0.01\pm0.03$ &$ 0.01\pm0.03$ \\ 
$0.25$ &$0.77$ &$0.46$ &$0.83$ &$ 0.83\pm0.03$ &$ 0.14\pm0.03$ &$ 0.00\pm0.03$ &$ 0.01\pm0.03$ &$ 0.02\pm0.03$ \\
\hline
\end{longtable*}
\end{ruledtabular}
\end{center}

\begin{center}
\begin{longtable*}{ccccc ccrr}
\caption{\label{tab:finalz} Final asymmetries and central values for the individual bins in the 9x9 $z_1\times z_2$ binning ($z_1,z_2 \in [0.2,0.275,0.35,0.425,0.5,0.575,0.65,0.725,0.825,1.0]$ and $m_1,m_2 \in [0.25,2.0]$ GeV/c$^2$) in \%. The following abbreviations are used: $s^2 = \sin^2\theta_t$, $1+c^2=1+\cos^2\theta_t$, $s\theta_1=\sin \theta_1 $ and $s\theta_2= \sin \theta_2$. }\\
\hline $\langle z_1 \rangle $ & $\langle z_2 \rangle $ & $\langle m_1 \rangle $ & $\langle m_2 \rangle $ & $ \langle s^2 \rangle/\langle 1+c^2 \rangle $ & $\langle s\theta_1 \rangle $ & $\langle s\theta_2 \rangle$ &  $a_{12}$ & $a_{12R}$ \\ \hline
\endhead
\hline\multicolumn{9}{c}{continued on next page}
\endfoot
\hline \hline
\endlastfoot
$0.25$& $0.25$& $0.46$& $0.46$& $0.86/1.14$& $0.98$& $0.98$&   $-0.48\pm 0.15\pm 0.12$&   $-0.47\pm 0.14\pm 0.12$ \\
$0.25$& $0.31$& $0.46$& $0.53$& $0.86/1.14$& $0.98$& $0.95$&   $-0.86\pm 0.11\pm 0.16$&   $-0.80\pm 0.10\pm 0.12$ \\
$0.25$& $0.39$& $0.46$& $0.59$& $0.86/1.14$& $0.98$& $0.93$&   $-1.21\pm 0.11\pm 0.14$&   $-1.08\pm 0.11\pm 0.14$ \\
$0.25$& $0.46$& $0.46$& $0.65$& $0.86/1.14$& $0.98$& $0.91$&   $-1.24\pm 0.13\pm 0.26$&   $-0.97\pm 0.12\pm 0.25$ \\
$0.25$& $0.53$& $0.46$& $0.70$& $0.85/1.15$& $0.98$& $0.90$&   $-1.71\pm 0.16\pm 0.29$&   $-1.49\pm 0.14\pm 0.27$ \\
$0.25$& $0.61$& $0.46$& $0.76$& $0.85/1.15$& $0.98$& $0.89$&   $-1.81\pm 0.19\pm 0.25$&   $-1.79\pm 0.18\pm 0.25$ \\
$0.25$& $0.68$& $0.46$& $0.81$& $0.85/1.15$& $0.98$& $0.88$&   $-2.20\pm 0.25\pm 0.32$&   $-2.03\pm 0.23\pm 0.37$ \\
$0.25$& $0.77$& $0.46$& $0.88$& $0.85/1.15$& $0.98$& $0.87$&   $-2.40\pm 0.31\pm 0.41$&   $-2.12\pm 0.29\pm 0.39$ \\
\end{longtable*}
\end{center}

\begin{center}
 \begin{longtable*}{rrrrrrrrr}
\caption{\label{tab:bgz} relative yields of uds, charm mixed, charged and tau contributions in the data sample for the individual bins in the 8x8 $m_1\times m_2$ binning ($m_1,m_2 \in [0.25,0.4,0.5,0.62,0.77,0.9,1.1,1.5,2.0 ]$ GeV/c$^2$ and $z_1,z_2 \in [0.2,1.0]$).}\\
$\langle z_1 \rangle $ & $\langle z_2 \rangle $ & $\langle m_1 \rangle $ & $\langle m_2 \rangle $ & uds & charm & mixed & charged & $\tau$ \\  \hline
\endhead
\hline\multicolumn{9}{c}{continued on next page}
\endfoot
\hline \hline
\endlastfoot
$0.37$ &$0.37$ &$0.35$ &$0.35$ &$ 0.59\pm0.03$ &$ 0.38\pm0.03$ &$ 0.01\pm0.03$ &$ 0.01\pm0.03$ &$ 0.01\pm0.03$ \\ 
$0.37$ &$0.39$ &$0.35$ &$0.45$ &$ 0.61\pm0.03$ &$ 0.36\pm0.03$ &$ 0.01\pm0.03$ &$ 0.01\pm0.03$ &$ 0.01\pm0.03$ \\ 
$0.37$ &$0.41$ &$0.35$ &$0.56$ &$ 0.62\pm0.03$ &$ 0.36\pm0.03$ &$ 0.01\pm0.03$ &$ 0.01\pm0.03$ &$ 0.01\pm0.03$ \\ 
$0.37$ &$0.45$ &$0.35$ &$0.69$ &$ 0.64\pm0.03$ &$ 0.33\pm0.03$ &$ 0.01\pm0.03$ &$ 0.01\pm0.03$ &$ 0.01\pm0.03$ \\ 
$0.37$ &$0.50$ &$0.35$ &$0.82$ &$ 0.67\pm0.03$ &$ 0.30\pm0.03$ &$ 0.01\pm0.03$ &$ 0.01\pm0.03$ &$ 0.01\pm0.03$ \\ 
$0.37$ &$0.54$ &$0.35$ &$0.98$ &$ 0.67\pm0.03$ &$ 0.30\pm0.03$ &$ 0.01\pm0.03$ &$ 0.01\pm0.03$ &$ 0.01\pm0.03$ \\ 
$0.37$ &$0.62$ &$0.35$ &$1.23$ &$ 0.62\pm0.03$ &$ 0.35\pm0.03$ &$ 0.01\pm0.03$ &$ 0.02\pm0.03$ &$ 0.01\pm0.03$ \\ 
$0.39$ &$0.72$ &$0.35$ &$1.62$ &$ 0.45\pm0.03$ &$ 0.51\pm0.04$ &$ 0.01\pm0.03$ &$ 0.02\pm0.03$ &$ 0.00\pm0.03$ \\
\hline
\end{longtable*}
\end{center}

\begin{center}
\begin{longtable*}{ccccc c  ccrr}
\caption{\label{tab:finalm} Final asymmetries and central values for the individual bins in the 8x8 $m_1\times m_2$ binning ($m_1,m_2 \in [0.25,0.4,0.5,0.62,0.77,0.9,1.1,1.5,2.0 ]$ GeV/c$^2$ and $z_1,z_2 \in [0.2,1.0]$) in \%. The following abbreviations are used: $s^2 = \sin^2\theta_t$, $1+c^2=1+\cos^2\theta_t$, $s\theta_1=\sin \theta_1 $ and $s\theta_2= \sin \theta_2$. }\\
\hline $\langle z_1 \rangle $ & $\langle z_2 \rangle $ & $\langle m_1 \rangle $ & $\langle m_2 \rangle $ & $\langle s^2 \rangle/ \langle 1+c^2 \rangle $ & $\langle s\theta_1 \rangle $ & $\langle s\theta_2 \rangle$ &  $a_{12}$ & $a_{12R}$ \\ \hline
\endhead
\hline\multicolumn{9}{c}{continued on next page}
\endfoot
\hline \hline
\endlastfoot
$0.36$& $0.36$& $0.35$& $0.35$& $0.83/1.17$& $0.89$& $0.89$&   $-0.03\pm 0.10\pm 0.20$&   $-0.05\pm 0.09\pm 0.16$ \\
$0.36$& $0.39$& $0.35$& $0.45$& $0.84/1.16$& $0.89$& $0.93$&   $ 0.16\pm 0.10\pm 0.39$&   $ 0.13\pm 0.09\pm 0.29$ \\
$0.36$& $0.40$& $0.35$& $0.56$& $0.84/1.16$& $0.89$& $0.94$&   $-0.70\pm 0.09\pm 0.14$&   $-0.63\pm 0.09\pm 0.13$ \\
$0.36$& $0.44$& $0.35$& $0.69$& $0.85/1.15$& $0.89$& $0.93$&   $-0.59\pm 0.10\pm 0.17$&   $-0.53\pm 0.09\pm 0.19$ \\
$0.37$& $0.49$& $0.35$& $0.83$& $0.85/1.15$& $0.89$& $0.93$&   $-1.18\pm 0.13\pm 0.21$&   $-1.01\pm 0.12\pm 0.19$ \\
$0.37$& $0.54$& $0.35$& $0.98$& $0.85/1.15$& $0.89$& $0.93$&   $-1.15\pm 0.16\pm 0.21$&   $-1.14\pm 0.15\pm 0.24$ \\
$0.37$& $0.63$& $0.35$& $1.24$& $0.85/1.15$& $0.89$& $0.94$&   $-1.13\pm 0.22\pm 0.26$&   $-1.02\pm 0.21\pm 0.21$ \\
$0.36$& $0.75$& $0.35$& $1.66$& $0.86/1.14$& $0.89$& $0.96$&   $-2.76\pm 0.59\pm 0.62$&   $-2.04\pm 0.55\pm 0.54$ \\
\hline
\end{longtable*}
\end{center}

\begin{center}
 \begin{longtable*}{rrrrrrrrr}
\caption{\label{tab:bgz} relative yields of uds, charm mixed, charged and tau contributions in the data sample for the the individual bins in the 8x8 $z_1\times m_1$ binning ($m_1 \in [0.25,0.4,0.5,0.62,0.77,0.9,1.1,1.5,2.0 ]$ GeV/c$^2$, $z_2 \in [0.2,0.27,0.33,0.4,0.5,0.6,0.7,0.8,1.0]$, $z_2 \in [0.2,1.0]$ and $m_2 \in [0.25,2.0]$ GeV/c$^2$).  Empty bins at high mass, low $z$ were removed.}\\
$\langle z_1 \rangle $ & $\langle z_2 \rangle $ & $\langle m_1 \rangle $ & $\langle m_2 \rangle $ & uds & charm & mixed & charged & $\tau$ \\  \hline
\endhead
\hline\multicolumn{9}{c}{continued on next page}
\endfoot
\hline \hline
\endlastfoot
$0.24$ &$0.24$ &$0.34$ &$0.60$ &$ 0.56\pm0.03$ &$ 0.41\pm0.03$ &$ 0.01\pm0.03$ &$ 0.02\pm0.03$ &$ 0.00\pm0.03$ \\ 
$0.30$ &$0.24$ &$0.35$ &$0.60$ &$ 0.59\pm0.03$ &$ 0.38\pm0.03$ &$ 0.01\pm0.03$ &$ 0.01\pm0.03$ &$ 0.01\pm0.03$ \\ 
$0.36$ &$0.24$ &$0.35$ &$0.60$ &$ 0.62\pm0.03$ &$ 0.35\pm0.03$ &$ 0.01\pm0.03$ &$ 0.01\pm0.03$ &$ 0.01\pm0.03$ \\ 
$0.44$ &$0.25$ &$0.35$ &$0.60$ &$ 0.67\pm0.03$ &$ 0.31\pm0.03$ &$ 0.00\pm0.03$ &$ 0.01\pm0.03$ &$ 0.01\pm0.03$ \\ 
$0.54$ &$0.24$ &$0.35$ &$0.60$ &$ 0.73\pm0.03$ &$ 0.25\pm0.03$ &$ 0.00\pm0.03$ &$ 0.01\pm0.03$ &$ 0.02\pm0.03$ \\ 
$0.64$ &$0.25$ &$0.35$ &$0.61$ &$ 0.78\pm0.03$ &$ 0.19\pm0.03$ &$ 0.00\pm0.03$ &$ 0.00\pm0.03$ &$ 0.03\pm0.03$ \\ 
$0.74$ &$0.24$ &$0.35$ &$0.61$ &$ 0.84\pm0.03$ &$ 0.11\pm0.03$ &$ 0.00\pm0.03$ &$ 0.00\pm0.03$ &$ 0.05\pm0.03$ \\ 
$0.84$ &$0.24$ &$0.36$ &$0.61$ &$ 0.85\pm0.04$ &$ 0.05\pm0.02$ &$ 0.00\pm0.03$ &$ 0.00\pm0.03$ &$ 0.10\pm0.03$ \\
\hline
\end{longtable*}
\end{center}

\begin{center}
\begin{longtable*}{cccccc ccrr}
\caption{\label{tab:finalmz} Final asymmetries and central values for the individual bins in the 8x8 $z_1\times m_1$ binning ($m_1 \in [0.25,0.4,0.5,0.62,0.77,0.9,1.1,1.5,2.0 ]$ GeV/c$^2$, $z_2 \in [0.2,0.27,0.33,0.4,0.5,0.6,0.7,0.8,1.0]$, $z_2 \in [0.2,1.0]$ and $m_2 \in [0.25,2.0]$ GeV/c$^2$) in \%. The following abbreviations are used: $s^2 = \sin^2\theta_t$, $1+c^2=1+\cos^2\theta_t$, $s\theta_1=\sin \theta_1 $ and $s\theta_2= \sin \theta_2$. Empty bins at high mass, low $z$ were removed.}\\
\hline $\langle z_1 \rangle $ & $\langle z_2 \rangle $ & $\langle m_1 \rangle $ & $\langle m_2 \rangle $ & $\langle s^2 \rangle / \langle 1+c^2 \rangle $ & $\langle s\theta_1 \rangle $ & $\langle s\theta_2 \rangle$ &  $a_{12}$ & $a_{12R}$ \\  \hline
\endhead
\hline\multicolumn{9}{c}{continued on next page}
\endfoot
\hline \hline
\endlastfoot
 $0.24$&$0.43$& $0.34$& $0.62$& $0.85/1.15$& $0.93$& $0.97$&   $-0.41\pm 0.09\pm 0.28$&   $-0.38\pm 0.09\pm 0.27$ \\
 $0.30$&$0.43$& $0.34$& $0.62$& $0.84/1.16$& $0.93$& $0.92$&   $-0.53\pm 0.08\pm 0.31$&   $-0.39\pm 0.08\pm 0.27$ \\
 $0.36$&$0.43$& $0.35$& $0.62$& $0.84/1.16$& $0.93$& $0.87$&   $-0.56\pm 0.09\pm 0.30$&   $-0.42\pm 0.08\pm 0.27$ \\
 $0.44$&$0.43$& $0.35$& $0.62$& $0.84/1.16$& $0.92$& $0.83$&   $-0.39\pm 0.10\pm 0.29$&   $-0.25\pm 0.10\pm 0.28$ \\
 $0.54$&$0.43$& $0.35$& $0.62$& $0.84/1.16$& $0.92$& $0.80$&   $-0.54\pm 0.15\pm 0.29$&   $-0.42\pm 0.14\pm 0.28$ \\
 $0.64$&$0.43$& $0.35$& $0.62$& $0.83/1.17$& $0.92$& $0.79$&   $-0.60\pm 0.24\pm 0.33$&   $-0.35\pm 0.22\pm 0.32$ \\
 $0.74$&$0.44$& $0.35$& $0.62$& $0.83/1.17$& $0.92$& $0.80$&   $-0.18\pm 0.40\pm 0.36$&   $-0.16\pm 0.37\pm 0.37$ \\
 $0.85$&$0.45$& $0.35$& $0.63$& $0.83/1.17$& $0.92$& $0.81$&   $-1.09\pm 0.69\pm 0.64$&   $-0.32\pm 0.64\pm 0.76$ \\
\hline
\end{longtable*}
\end{center}

\begin{figure}[h]
\begin{center}
\includegraphics[width=0.8\textwidth]{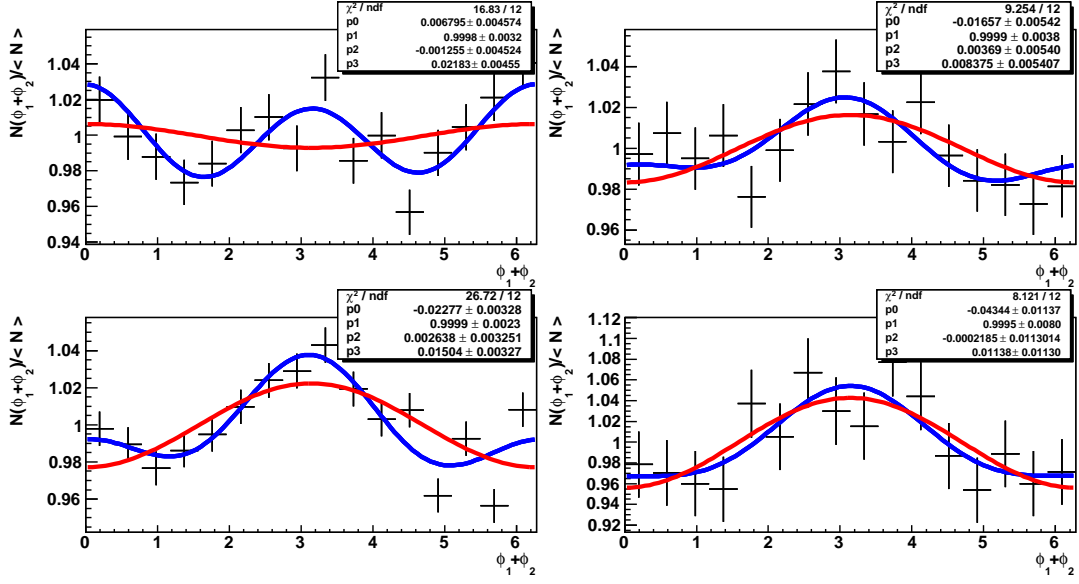}
\caption{\label{fig:asydists} Normalized azimuthal yields for the $z_1,z_2$ bins ([0.2,0.275], [0.2,0.275]); ([0.275,0.35],[0.65,0.725]); ([0.425,0.5],[0.425,0.5]) and ([0.65,0.725],[0.265,0.725]) are displayed as a function of the azimuthal angular combination $\phi_1+\phi2$ for one Belle running period. The fit results of $p0 \cos(\phi_1+\phi_2) + p1$ (red) and $p0 \cos(\phi_1+\phi_2) + p1 + p2 \sin(\phi_1+\phi_2) + p3 \cos(2(\phi_1+\phi_2)) $ (blue) are also displayed. }
\end{center}
\end{figure}
\cleardoublepage

\end{document}